\documentclass[12pt]{article}
\usepackage{amsmath} 
\usepackage{longtable}
\usepackage{graphicx,epsfig}    
\usepackage{color}
\usepackage{cite}
\usepackage{amssymb}
\newcommand\as{\alpha_{\mathrm{S}}} 
\newcommand\f[2]{\frac{#1}{#2}}

\def\beq{\begin{equation}} 
\def\eeq{\end{equation}} 
\def\to{\rightarrow} 
\def\nn{\nonumber}

\def\b0{\beta_0}

\def\GE{\gamma_E}
\def\beeq{\begin{eqnarray}}
\def\eeeq{\end{eqnarray}}

\def\mur{\mu_R} 
\def\muf{\mu_F}
\def\mur2{\mu_R^2} 
\def\muf2{\mu_F^2}

\newcommand {\apgt} {\ {\raise-.5ex\hbox{$\buildrel>\over\sim$}}\ }
\newcommand {\aplt} {\ {\raise-.5ex\hbox{$\buildrel<\over\sim$}}\ }

\begin{document} 

\setlength{\parskip}{0.15cm}
\setlength{\baselineskip}{0.52cm}

\begin{titlepage}
\renewcommand{\thefootnote}{\fnsymbol{footnote}}
\thispagestyle{empty}
\noindent
\vspace{1.0cm}

\begin{center}
{\bf \Large 
Higgs pair production at next-to-next-to-leading \\
\vspace{0.3cm}
 logarithmic accuracy at the LHC}\\
  \vspace{1.25cm}
{\large
Daniel de Florian$\,$\footnote{deflo@df.uba.ar} and 
Javier Mazzitelli$\,$\footnote{jmazzi@df.uba.ar} \\
}
 \vspace{1.25cm}
 {\it
    Departamento de F\'\i sica, FCEyN, Universidad de Buenos Aires \\
   (1428) Pabell\'on 1, Ciudad Universitaria, Capital Federal, Argentina \\
 }
  \vspace{1.5cm}
  \large {\bf Abstract}
  \vspace{-0.2cm}
\end{center}
We perform the threshold resummation for Higgs pair production in the dominant gluon fusion channel to 
next-to-next-to-leading logarithmic (NNLL) accuracy.

The calculation includes the matching to the next-to-next-to-leading order (NNLO) cross section obtained in the heavy top-quark limit, and results in an increase of the inclusive cross section up to $7\%$ at the LHC with centre-of-mass energy $E_{cm}=14\text{ TeV}$,
for the choice of factorization and renormalization scales $\mu_F=\mu_R=Q$, being $Q$ the invariant mass of the Higgs pair system.

After the resummation is implemented, we estimate the theoretical uncertainty from the perturbative expansion to be reduced to about $\pm5.5\%$, plus $\sim 10\%$ from finite top-mass effects.
The resummed cross section turns out to be rather independent of the value chosen for the central factorization and renormalization scales in the usual range $(Q/2,Q)$.

\hfill

\end{titlepage}
\setcounter{footnote}{1}
\renewcommand{\thefootnote}{\fnsymbol{footnote}}

%
\section{Introduction}

The experimental collaborations ATLAS and CMS have discovered in 2012 a new particle at the LHC \cite{Aad:2012tfa,Chatrchyan:2012ufa}, whose properties are so far compatible with the Standard Model (SM) Higgs boson.
In order to be able to distinguish between the SM and other new physics scenarios, it is crucial to measure its properties as accurate as possible.
In particular, to understand the relation between the discovered boson and the electroweak symmetry breaking mechanism it is essential to determine its couplings to fermions and gauge bosons, and verify its proportionality to  the particle masses.
Furthermore, it is necessary to measure the Higgs self-interactions, in order to start reconstructing the scalar potential, which is responsible for the spontaneous symmetry breaking.
While the Higgs quartic coupling is currently out of reach \cite{Plehn:2005nk}, several studies have recently shown that a measurement of the Higgs self-coupling can be achieved at a luminosity upgraded LHC via Higgs pair production \cite{Baur:2002qd,
Dolan:2012rv,Papaefstathiou:2012qe,
Baglio:2012np,Baur:2003gp,Goertz:2013kp,Gouzevitch:2013qca,Bhattacherjee:2014bca}.
This could be one of the main goals of the high-luminosity run of the LHC.

As it happens for single Higgs production, the SM Higgs pair production is dominated by the gluon fusion mechanism mediated by a heavy-quark loop.
At leading order (LO) in QCD perturbation theory \cite{Glover:1987nx,Eboli:1987dy,Plehn:1996wb}, this process can occur either via a box diagram, $gg\to HH$, or a triangle diagram, $gg\to H^*\to HH$, being the latter the only one sensitive to the Higgs trilinear coupling.
Given that this process is already one-loop level at LO, higher order corrections are very difficult to compute in the full theory.

The next-to-leading order (NLO) corrections have been computed in the large top-mass approximation in Ref. \cite{Dawson:1998py}, and more recently also the next-to-next-to-leading order (NNLO) cross section became available within the same effective theory \cite{deFlorian:2013jea}.
The QCD corrections were found to be large, resulting in about a $100\%$ increase from LO to NLO, and a still sizeable $20\%$ increment from NLO to NNLO at a collider center of mass energy of $14\text{ TeV}$.
The higher order corrections are almost completely dominated by soft and virtual terms, as it was shown in Ref. \cite{deFlorian:2013jea}.
The theoretical uncertainty arising from missing higher orders in the perturbative expansion was estimated to be about $\pm 8.5\%$ at NNLO for this energy. On top of this, one should add the uncertainties of the strong coupling and parton flux determination, plus the unknown finite quark-mass effects.
In order to reduce these uncertainties, one should compute higher order corrections and finite top-mass effects, respectively. 

In this work, we compute the dominant effect of the uncalculated higher-order terms by exploiting the resummation of soft-gluon emission, working within the large top-mass approximation.
We provide numerical results for the LHC up to the next-to-next-to-leading logarithmic (NNLL) accuracy.
Our calculation consistently includes the matching to the NNLO cross section.

The paper is organized as follows. In Section \ref{results} we define our notation and present all the expressions needed to perform threshold resummation for Higgs pair production up to NNLL.
In Section \ref{phenomenology} we present the numerical results for the LHC, comparing the fixed order and the resummed cross sections in several aspects.
Finally, in Section \ref{conclusions} we present our conclusions.

%
\section{Threshold resummation for Higgs pair production}\label{results}

We consider the production of a Higgs boson pair via top-quark loops.
We work within the large top-mass approximation, where the effective gluon-Higgs coupling is given by the following Lagrangian
\beq
{\cal L}_{\text{eff}}=-\f{1}{4}G_{\mu\nu}G^{\mu\nu}
\left(
C_H\f{H}{v}-C_{HH}\f{H^2}{v^2}
\right),
\eeq
where $v\simeq 246\text{ GeV}$ is the Higgs vacuum expectation value and $G_{\mu\nu}$ stands for the gluonic field strength tensor.
The perturbative expansions of $C_H$ \cite{Chetyrkin:1997iv,Chetyrkin:2005ia,Kramer:1996iq,Schroder:2005hy} and $C_{HH}$ \cite{Djouadi:1991tka,Grigo:2014jma} are known up to the order needed for our calculation, i.e. ${\cal O}\left(\as^3\right)$.

We start by setting the notation for the fixed order calculation. The hadronic cross section for a center-of-mass energy of the collider $s_H$, differential in the Higgs pair invariant mass $Q$, can be written in the following way
\begin{align}
\label{had}
Q^2\,\f{d\sigma}{dQ^2}(s_H,Q^2) \equiv \sigma(s_H,Q^2) =& 
\sum_{a,b} \int_0^1 dx_1 \;dx_2 \; f_{a/h_1}(x_1,\mu_F^2) 
\;f_{b/h_2}(x_2,\mu_F^2) 
\nn \\
& \times
\int_0^1 dz \;\delta\!\left(z -
\frac{\tau}{x_1x_2}\right) 
 \hat{\sigma}_0\,z\;G_{ab}(z;\as(\mu_R^2), Q^2/\mu_R^2;Q^2/\mu_F^2) \;,
\end{align}
where $\tau=Q^2/s_H$, $\mu_R$ and $\mu_F$ are the renormalization and factorization scales respectively, and $\hat{\sigma}_0$ is the Born level partonic cross section. 
The parton densities of the colliding hadrons are denoted by 
$f_{a/h}(x,\mu_F^2)$ and the subscripts $a,b$ label the type
of massless partons ($a,b=g,q_f,{\bar q}_f$,
with $N_f=5$ different flavours of light quarks).
The hard coefficient function $G_{ab}$ can be expanded in terms of powers of the QCD renormalized coupling $\as(\mu_R^2)$ (in the following, the dependence of $\as$ on $\mu_R$ is understood):
\begin{align}
\label{expansion}
G_{ab}(z;\as, Q^2/\mu_R^2;Q^2/\mu_F^2) &=
\sum_{n=0}^{+\infty} \left(\f{\as}{2\pi}\right)^n
\;G_{ab}^{(n)}(z;Q^2/\mu_R^2;Q^2/\mu_F^2)\;.
\end{align}
We use the $\overline{\text{MS}}$ scheme for the renormalization of the strong coupling.

Given that the soft-gluon resummation has to be carried out in Mellin (or $N$-moment) space \cite{Sterman:1986aj,Catani:1989ne}, we introduce now the corresponding notation.
We consider the Mellin transform $\sigma_N(Q^2)$ of the hadronic cross
section. The $N$-moments with respect to $\tau=Q^2/s_H$
at fixed $Q$ are defined as follows:
\begin{equation}
\label{sigman}
\sigma_N(Q^2) \equiv \int_0^1 \;d\tau \;\tau^{N-1} \;
\sigma (s_H,Q^2) 
\;\;.
\end{equation} 
In $N$-moment space, Eq.~(\ref{had}) takes the following simple factorized form
\begin{equation}
\label{hadn}
\sigma_{N-1}(Q^2) = \hat{\sigma}_0 \;\sum_{a,b}
\; f_{a/h_1, \, N}(\mu_F^2) \; f_{b/h_2\, N}(\mu_F^2) 
\; {G}_{ab,\, N}(\as, Q^2/\mu_R^2;Q^2/\mu_F^2) \;,
\end{equation}
where we have introduced the $N$-moments  of the
parton distributions and the hard coefficient function as
\begin{align} 
\label{pdfn}
f_{a/h, \, N}(\mu_F^2) &= \int_0^1 \;dx \;x^{N-1} \;
f_{a/h}(x,\mu_F^2) \;, \\
\label{gndef}
G_{ab,\, N} &= \int_0^1 dz \;z^{N-1} \;G_{ab}(z) \;\;.
\end{align}
Once these $N$-moments are known,
the physical cross section in $z$-space can be obtained by Mellin inversion by
\begin{align}
\sigma^{res} (s_H,Q^2)
= \hat{\sigma}_0 \;\sum_{a,b} & 
\;\int_{C_{MP}-i\infty}^{C_{MP}+i\infty}
\;\frac{dN}{2\pi i} \;\left( \frac{Q^2}{s_H} \right)^{-N+1} \;
f_{a/h_1, \, N}(\mu_F^2) \; f_{b/h_2\, N}(\mu_F^2) \nonumber \\
\label{invmt}
& \times
\; {G}_{ab,\, N}(\as, Q^2/\mu_R^2;Q^2/\mu_F^2) \;,
\end{align} 
where the constant $C_{MP}$ that defines the integration contour in the $N$-plane
is on the right of all the possible singularities of the integrand, as defined in the Minimal Prescription introduced in \cite{Catani:1996yz}.

We want to consider the all-order summation of the enhanced threshold ($z\to 1$) contributions, which corresponds to the limit $N\to\infty$ in Mellin space.
Given that $gg$ is the only partonic channel which is not suppressed in this limit, we only need to consider its contribution.
The resummation of soft-gluon effects is achieved  by organizing the partonic coefficient function  in Mellin space as
\begin{align}
\label{resfdelta}
G_{{gg},\, N}^{{\rm (res)}}(\as, Q^2/\mu_R^2;Q^2/\mu_F^2) 
&=  
C_{gg}(\as,Q^2/\mu^2_R;Q^2/\mu_F^2) \nn \\ 
&\cdot  \Delta_{N}(\as,Q^2/\mu^2_R;Q^2/\mu_F^2) +
{\cal O}(1/N)\; , 
\end{align}
The large logarithmic corrections (that appear as $\as^n\ln^{2n-k} N$ in Mellin space) are exponentiated in the  Sudakov radiative factor $\Delta_N$, which depends only on the dynamics of soft gluon emission from the initial state partons.
It can be expanded as
\begin{align} 
\label{calgnnll} 
~\vspace{-.5cm} \ln \Delta_{N}  \!\left(\as,\ln N;\frac{Q^2}{\mu^2_R}, 
\frac{Q^2}{\mu_F^2}\right) &=
\ln N \; g^{(1)}(\beta_0 \as \ln N) + 
g^{(2)}(\beta_0 \as \ln N, Q^2/\mu^2_R;Q^2/\mu_F^2 )
\nonumber 
\\ 
&+ \as 
\;g^{(3)}(\beta_0 \as\ln N,Q^2/\mu^2_R;Q^2/\mu_F^2 ) 
\nonumber \\
&+ \sum_{n=4}^{+\infty}  \as^{n-2}
\; g^{(n)}(\beta_0 \as\ln N,Q^2/\mu^2_R;Q^2/\mu_F^2 )\;. 
\end{align}

The function
$\ln N \; g^{(1)}$ resums all the {\em leading} logarithmic (LL) contributions
$\as^n \ln^{n+1}N$, $g^{(2)}$ contains the {\em next-to-leading} logarithmic 
(NLL) terms $\as^n \ln^{n}N$, $\as\, g^{(3)}$ collects
the {\em next-to-next-to-leading} logarithmic (NNLL) terms 
$\as^{n+1} \ln^{n}N$, and so forth. All the perturbative coefficients required
to construct the $g^{(1)}, g^{(2)}$ and $g^{(3)}$ functions are known and
only depend on the nature of the initiating partons.
Their explicit expression can be found, for instance, in Refs.\cite{Catani:2003zt,Vogt:2000ci}.

The function $C_{gg}(\as)$ contains all the contributions that are 
constant in the large-$N$ limit. They are produced by the hard virtual 
contributions and non-logarithmic soft corrections, and  
can be computed as a power series expansion in $\as$: 
\begin{equation}
\label{Cfun}
C_{gg}(\as,Q^2/\mu^2_R;Q^2/\mu_F^2) =  
1 + \sum_{n=1}^{+\infty} \;  
\left( \frac{\as}{2\pi} \right)^n \; 
C_{gg}^{(n)}(Q^2/\mu^2_R;Q^2/\mu_F^2) \;\;.
\end{equation}

The $C_{gg}^{(i)}$ coefficient, needed to perform N$^i$LL resummation, can be obtained from the N$^i$LO fixed order computation.
The only process-dependent contribution to $C_{gg}^{(i)}$ arises from the virtual corrections, given that the soft contributions are universal.
In fact, in Ref. \cite{deFlorian:2012za} we derived a universal formula for the coefficients needed up to NNLL accuracy, only dependent on the infrared regulated one and two-loop amplitudes (more recently the calculation was extended to N$^3$LL in Ref. \cite{Catani:2014uta}).
Then, we can obtain the expressions for $C_{gg}^{(1)}$ and $C_{gg}^{(2)}$ from the explicit two-loop calculation performed in Ref. \cite{deFlorian:2013uza}.
Specifically, we have
\beeq
C_{gg}^{(1)}&=&
C_A \f{4\pi^2}{3} + 4 C_A \GE^2 + \frac{\hat{\sigma}^{(1)}_{\text{fin}}}{\hat{\sigma}_0}\;,
\\
C_{gg}^{(2)}&=&
C_A^2 \bigg(
-\frac{55 \zeta_3}{36}-14 \gamma_E  \zeta_3+\frac{607}{81}+\frac{404 \gamma_E
   }{27}+\frac{134 \gamma_E ^2}{9}+\frac{44 \gamma_E ^3}{9}+8\GE^4 \nn\\
   &+&\frac{67 \pi
   ^2}{16}+\frac{14 \gamma_E ^2 \pi ^2}{3}+\frac{91 \pi ^4}{144}
\bigg)
+C_A N_f \left(
\frac{5 \zeta_3}{18}-\frac{82}{81}-\frac{56 \gamma_E }{27}-\frac{20 \gamma_E
   ^2}{9}-\frac{8 \gamma_E ^3}{9}-\frac{5 \pi ^2}{8}
\right)\nn\\
&+& \beta_0^2 \f{11\pi^4}{3}
+C_A \frac{\hat{\sigma}^{(1)}_{\text{fin}}}{\hat{\sigma}_0}
\left(\f{4\pi^2}{3}+4\GE^2\right)+
\frac{\hat{\sigma}^{(2)}_{\text{fin}}}{\hat{\sigma}_0}\;,
\eeeq
where $\GE$ is the Euler number, $\zeta_n$ represents the Riemann zeta function and $\beta_0=(11C_A-2N_f)/12\pi$.
The infrared-regulated one and two-loop corrections $\hat\sigma^{(1)}_{\text{fin}}$ and $\hat\sigma^{(2)}_{\text{fin}}$ are defined by the following set of formulae (for $\mu_R=\mu_F=Q$)
\beeq
C_{LO}&=&\frac{3M_H^2}{Q^2-M_H^2+iM_H\Gamma_H}-1\,,
\\
\frac{\hat{\sigma}^{(1)}_{\text{fin}}}{\hat{\sigma}_0}
&=&
\f{1}{\left|C_{LO}\right|^2}\left(
11 \left|C_{LO}\right|^2 + \f{4}{3}\text{Re}\left(C_{LO}\right)
\right)\,,\\
\frac{\hat{\sigma}^{(2)}_{\text{fin}}}{\hat{\sigma}_0}
&=&
\f{1}{\left|C_{LO}\right|^2 (t_+ - t_-)}\int_{t_-}^{t_+}dt
\,\Big\{
\left|C_{LO}\right|^2 {\cal F}^{(2)}
+ \text{Re}\left(C_{LO}\right) {\cal R}^{(2)}
+ \text{Im}\left(C_{LO}\right) {\cal I}^{(2)}
+ {\cal V}^{(2)}
\Big\}\,,
\eeeq
with $t_{\pm}=-\tfrac{1}{2} \left(
Q^2-2M_H^2\mp Q\sqrt{Q^2-4M_H^2}
\right)\,$ and where for the sake of brevity we refer the expressions of ${\cal F}^{(2)}$, ${\cal R}^{(2)}$, ${\cal I}^{(2)}$ and ${\cal V}^{(2)}$ to Ref. \cite{deFlorian:2013uza} (supplemented with Ref. \cite{Grigo:2014jma} for $C_{HH}^{(2)}$).

Finally, in order to profit from the fixed order calculation we implement the corresponding matching.
As usual, we expand the resummed cross section to ${\cal O}(\alpha_s^4)$ \footnote{The lowest order cross section $\hat{\sigma}_0$ starts at ${\cal O}(\alpha_s^2)$}, subtract the expanded result from the resummed one, and add the full NNLO cross section, as
\begin{align}
\sigma^{NNLL} (s_H,Q^2) = \sigma^{res} (s_H,Q^2) - \sigma^{res} (s_H,Q^2)|_{{\cal O}(\alpha_s^4)} + \sigma^{NNLO} (s_H,Q^2) \, ,
\end{align} 
and similarly for LL and NLL.

With all the previous definitions and results, we are ready to perform the threshold resummation up to NNLL.
For more details of the resummation formalism, see for instance Ref. \cite{Catani:2003zt}.

\section{NNLL phenomenology}\label{phenomenology}

We present in this section the phenomenological results.
For the computation we take the Higgs mass to be $M_H=125\text{ GeV}$.
All the results are normalized by the exact LO top mass dependence, with $M_t=173.21\text{ GeV}$.
For the parton luminosities and strong coupling we use the MSTW2008 sets, consistently at each perturbative order (i.e.  LO PDFs and one-loop $\as$ evolution for LO and LL cross sections, etc.).
The scale uncertainty was evaluated by varying independently the renormalization and factorization scales in the range $\mu_0/2\leq\mu_R,\mu_F\leq 2\mu_0$ with the constraint $1/2\leq\mu_R/\mu_F<2$, where $\mu_0$ is the central scale.
The analysis was performed for two choices of the central scale: $\mu_0=Q$ and $\mu_0=Q/2$, being $Q$ the invariant mass of the Higgs pair system.

The contributions from all the relevant partonic channels are always included in our numerical results.
As described in the previous section, the threshold resummation only applies for the $gg$ channel. With the corresponding matching we also account for the other partonic subprocesses at the corresponding fixed order accuracy.

\begin{figure}[t!]
\begin{center}
\begin{tabular}{c c}
\includegraphics[width=8cm]{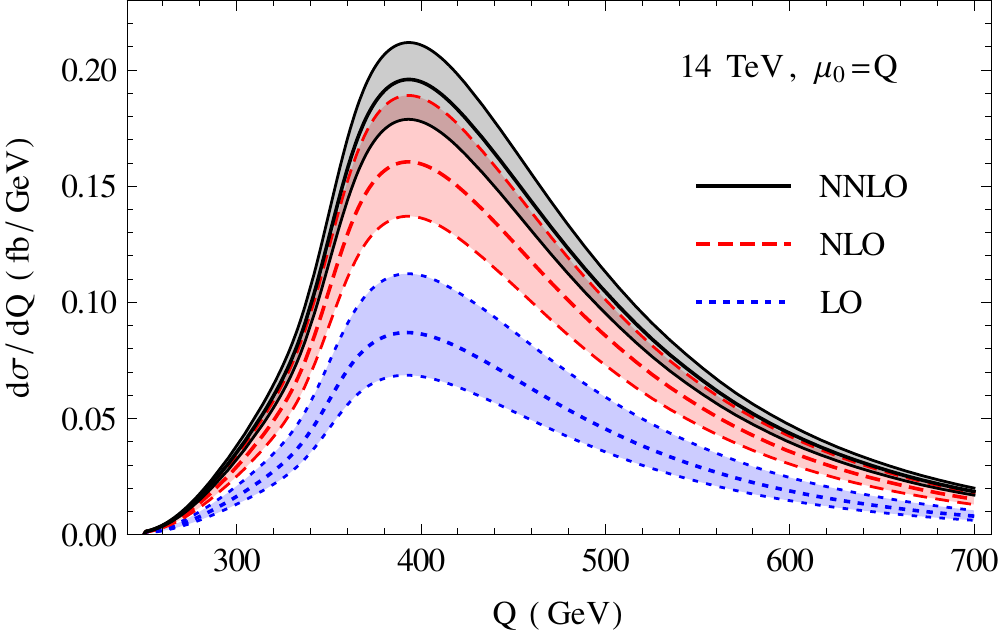}
&
\hspace{0.6cm}
\includegraphics[width=8cm]{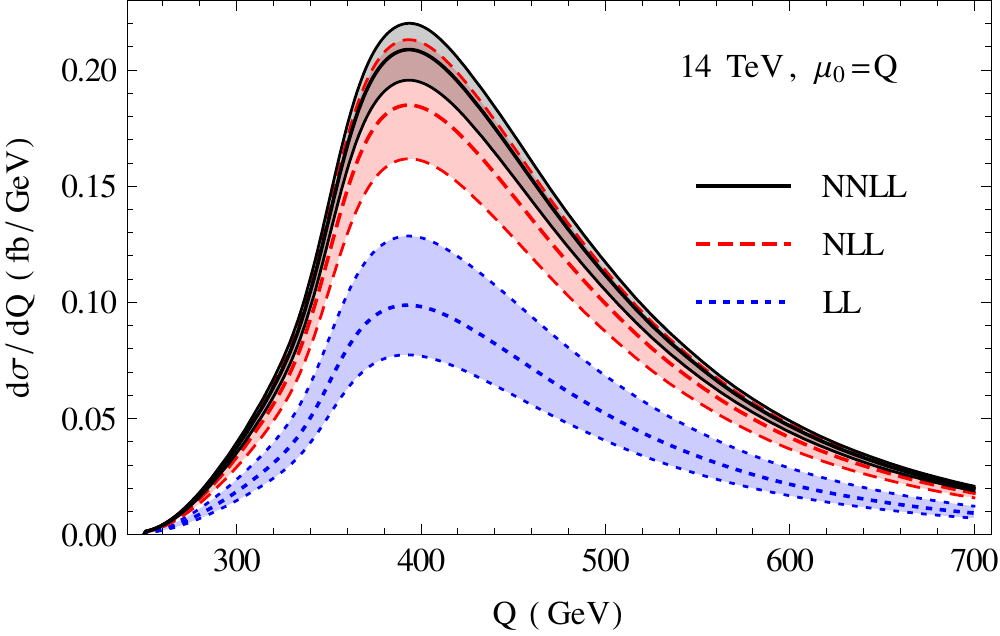}
\end{tabular}
\end{center}
\vspace{-0.7cm}
\caption{\label{dsdQ_14_mu1}\small
The Higgs pair invariant mass distribution for $E_{cm}=14\text{ TeV}$ and the central scale $\mu_0=Q$, for the fixed order (left) and resummed (right) predictions.
In the left (right) we show the LO (LL), NLO (NLL) and NNLO (NNLL) curves, with blue dotted, red dashed and black solid lines respectively.
}

\begin{center}
\begin{tabular}{c c}
\includegraphics[width=8cm]{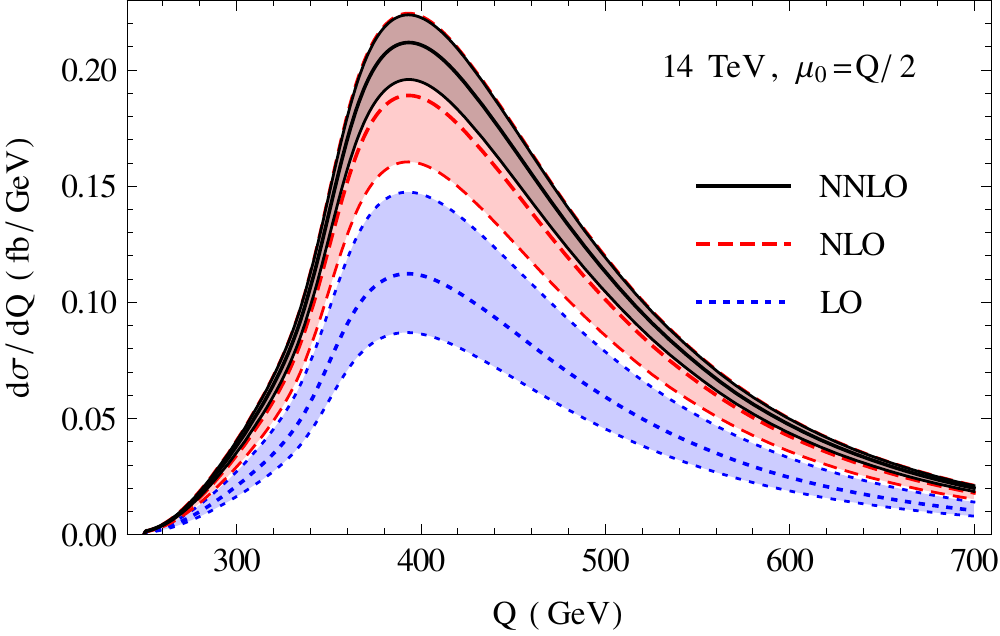}
&
\hspace{0.6cm}
\includegraphics[width=8cm]{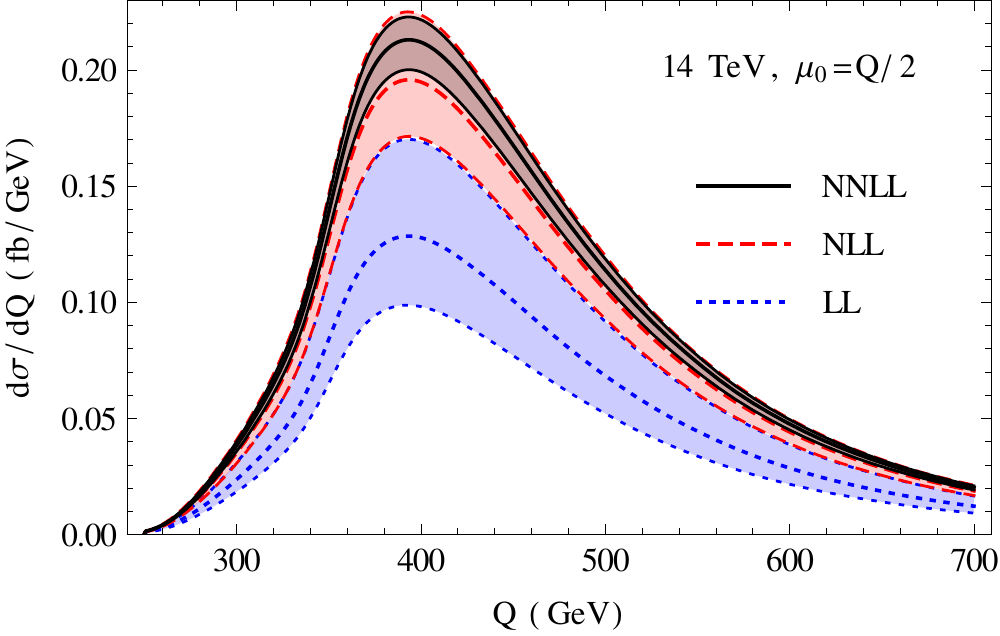}
\end{tabular}
\end{center}
\vspace{-0.7cm}
\caption{\label{dsdQ_14_mu05}\small
The Higgs pair invariant mass distribution for $E_{cm}=14\text{ TeV}$ and the central scale $\mu_0=Q/2$, for the fixed order (left) and resummed (right) predictions.
The color coding is the same of Figure \ref{dsdQ_14_mu1}.}
\end{figure}

We start by showing the Higgs pair invariant mass distribution for a collider center of mass energy $E_{cm}=14\text{ TeV}$.
In Figure \ref{dsdQ_14_mu1} we present the results corresponding to the central scale $\mu_0=Q$, while in Figure \ref{dsdQ_14_mu05} the ones corresponding to $\mu_0=Q/2$ are shown.
For both figures, in the left plot we present the fixed order prediction (at LO, NLO and NNLO) while in the right one we show the resummed cross section (at LL, NLL and NNLL).
\footnote{For simplicity, we always label our resummed predictions as LL, NLL and NNLL. As explained before, these results include the matching to the fixed order cross section, so they should be interpreted as LL+LO, NLL+NLO and NNLL+NNLO respectively.}

In the first place we can observe that, with the exception of the $\mu_0=Q/2$ resummed distributions,  there is no overlap between the LO (LL) and NLO (NLL) bands, and it is only at second order that a sensible superposition of the bands occurs. 
We can also see from the plots that at every order the inclusion of the resummed contributions results in an increase of the cross section. Also, we can observe that the size of the uncertainty band at NNLL is always smaller than the corresponding NNLO one. This effect is more clear with the choice $\mu_0=Q$, for which also a better overlap between the NNLL and NLL bands is observed, with respect to the NNLO and NLO ones.
The fixed order and resummed distributions have less differences for $\mu_0=Q/2$, as was already observed for single Higgs production, where the choice $\mu_0=M_H/2$ partially mimics some of the threshold resummation effects.
Regarding the shape of the distributions,
we observe very small differences after the resummation is performed.
This is due to the fact that the relative size of the resummed contributions has a rather small dependence on the Higgs pair invariant mass.

\begin{figure}[t!]
\begin{center}
\begin{tabular}{c c c c}
\hspace{0.0cm}
\includegraphics[height=8cm]{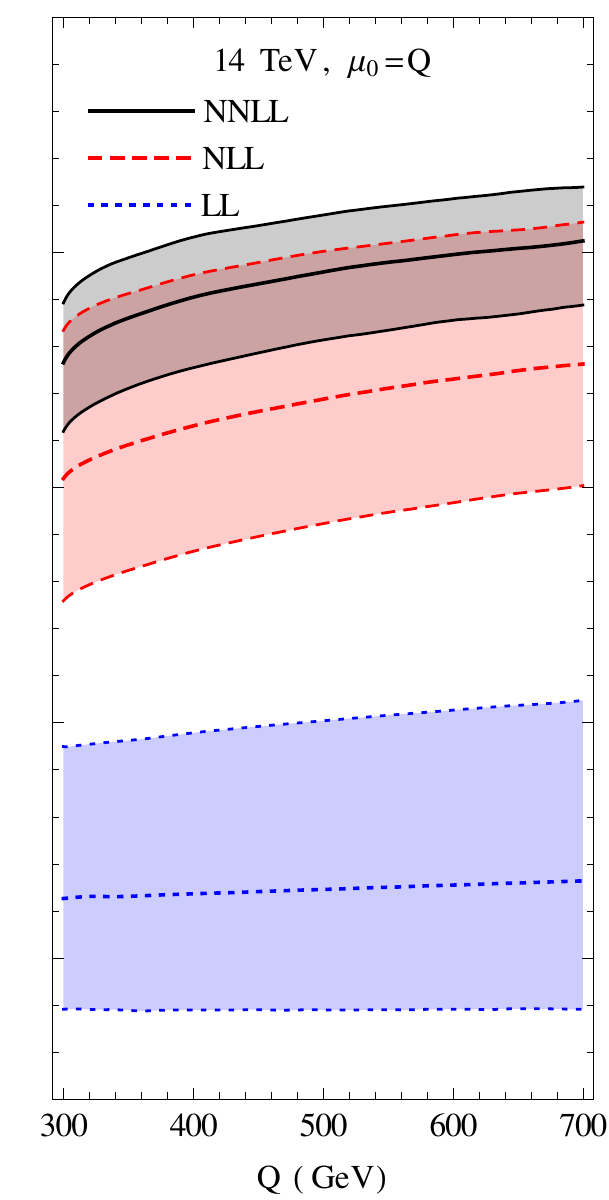}
&
\hspace{-8.9cm}
\includegraphics[height=8cm]{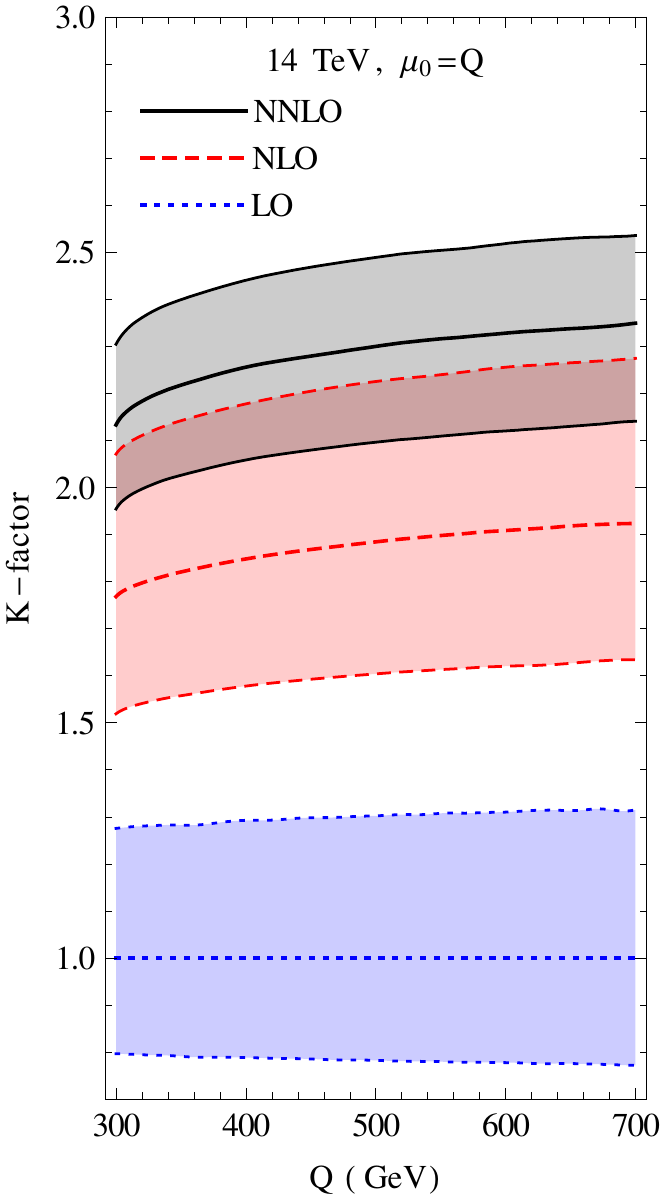}
&
\hspace{8.5cm}
\includegraphics[height=7.9cm]{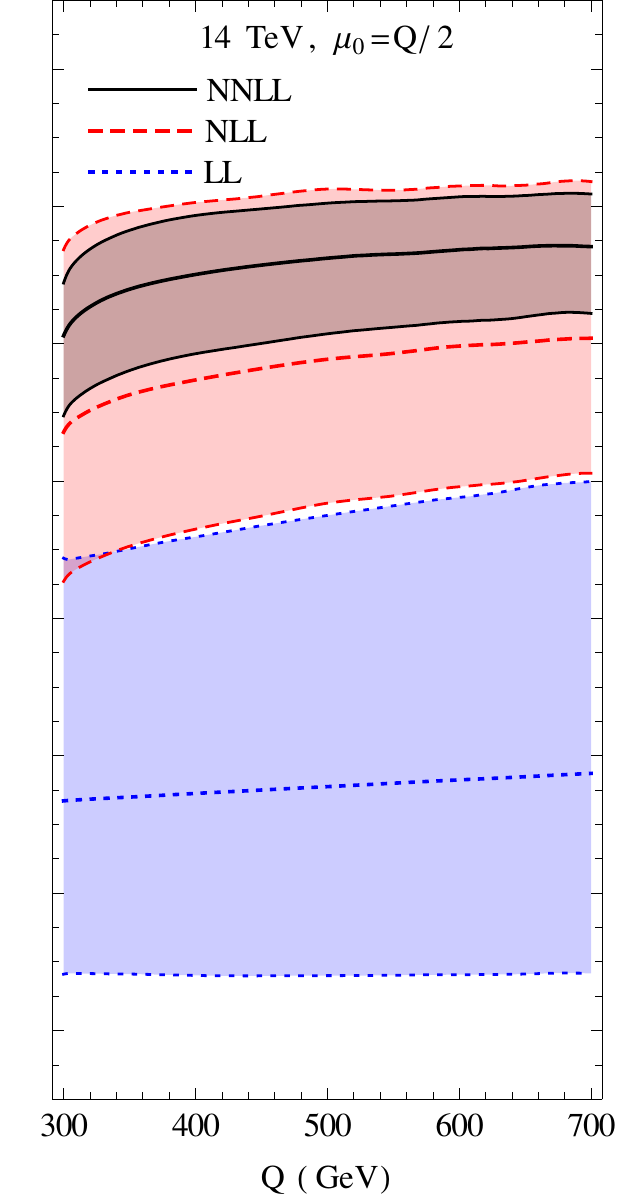}
&
\hspace{-8.9cm}
\includegraphics[height=7.9cm]{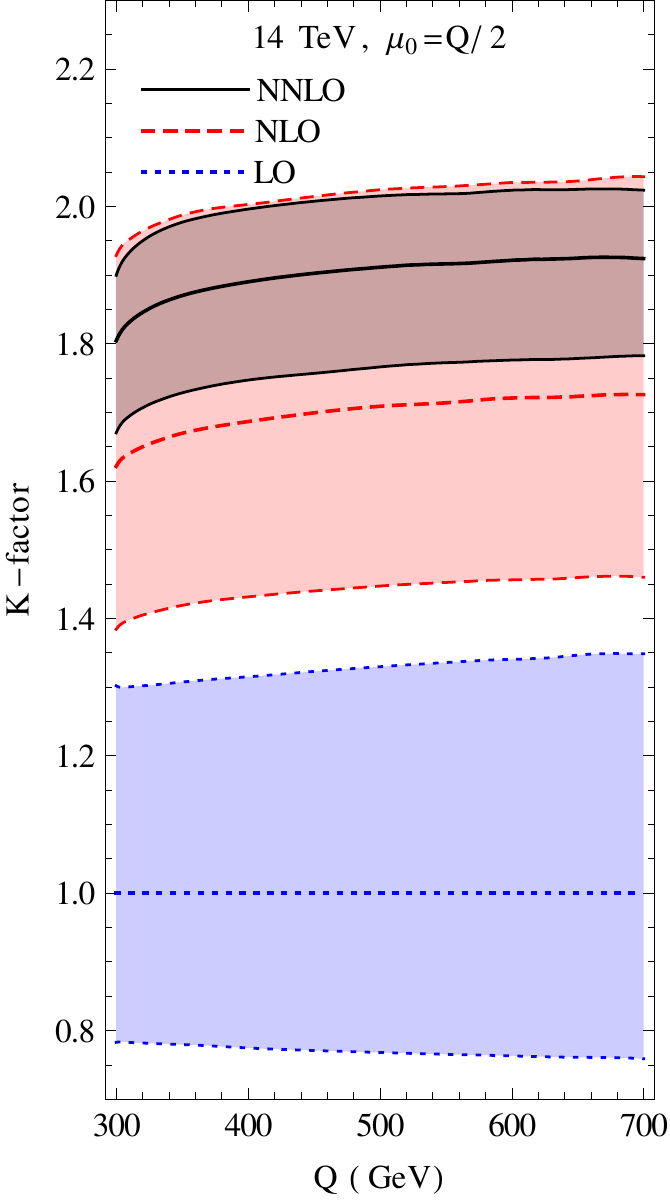}
\end{tabular}
\end{center}
\vspace{-0.7cm}
\caption{\label{Kfactors}\small
The $K$-factors for the fixed order and resummed cross sections as a function of the Higgs pair invariant mass, for $E_{cm}=14\text{ TeV}$.
The left (right) panel shows the results for $\mu_0=Q$ ($\mu_0=Q/2$).
The color coding is the same of Figure \ref{dsdQ_14_mu1}.
}
\end{figure}

In Figure \ref{Kfactors} we present the $K$-factors, defined as the ratio between a given prediction and the LO one. For the denominator we fix $\mu_R=\mu_F=\mu_0$.
We observe, in more detail, the same features described above at the level of the cross section. 
In particular, it is visible that the resummed series has a better convergence than the fixed order one, exhibiting a larger overlap between the first and second order bands.

\begin{figure}
\begin{center}
\begin{tabular}{c c}
\includegraphics[width=8cm]{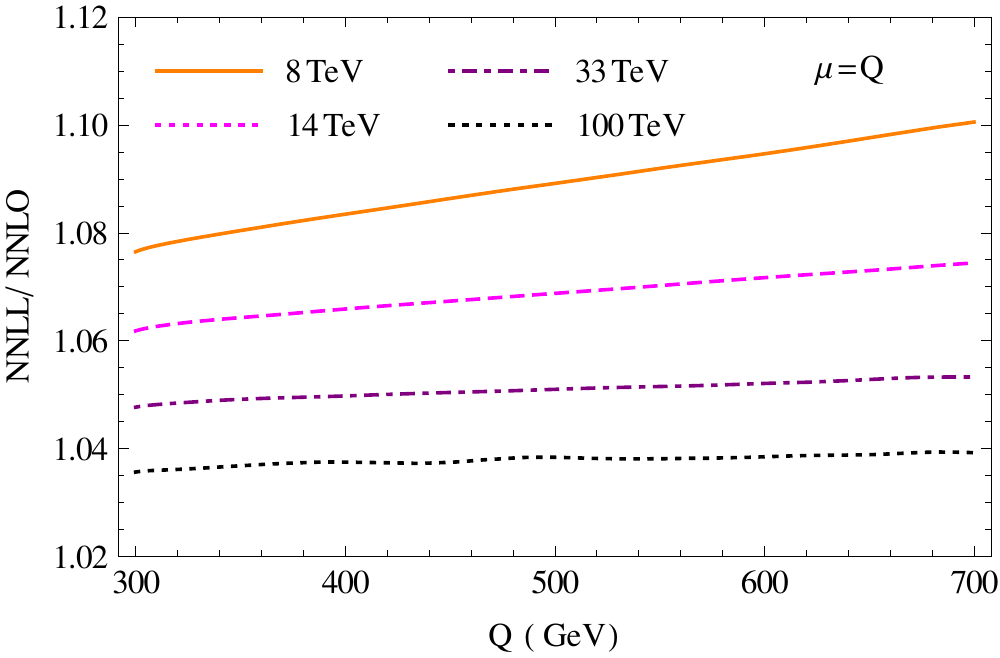}
&
\hspace{0.6cm}
\includegraphics[width=8cm]{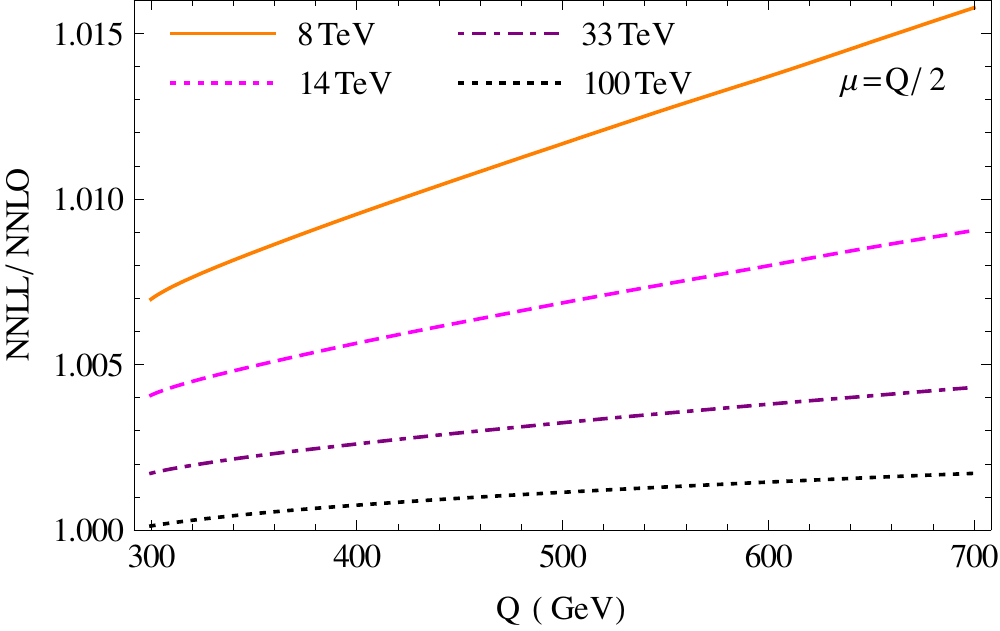}
\end{tabular}
\end{center}
\vspace{-0.7cm}
\caption{\label{ratio}\small
The ratio between the NNLL and the NNLO predictions as a function of the Higgs pair invariant mass, for the scales $\mu=Q$ (left) and $\mu=Q/2$ (right).
Results are shown for center of mass energies of $8\text{ TeV}$ (orange solid), $14\text{ TeV}$ (magenta dashed), $33\text{ TeV}$ (purple dot-dashed) and $100\text{ TeV}$ (black dotted).
}
\end{figure}

In Figure \ref{ratio} we show the ratio between the NNLL and the NNLO predictions, again as a function of the Higgs pair invariant mass, for different collider energies.
The ratio shows an almost linear dependence on $Q$, increasing for higher invariant masses.
Actually, this is expected because resummation contributions are enhanced when the process becomes closer to the partonic threshold.
The same feature is reflected by the fact that the resummation contributions are relatively smaller for larger collider energies.
We can also observe, as it was already clear from Figures \ref{dsdQ_14_mu1} and \ref{dsdQ_14_mu05}, that the ratio between NNLL and NNLO is significantly smaller for the scale choice $\mu_R=\mu_F=\mu=Q/2$.
At the total cross section level, for example, we find that the increase in the NNLL result with respect to the NNLO prediction is of $6.8\%$ for $E_{cm}=14\text{ TeV}$ and $\mu=Q$, while it drops down to $0.65\%$ for $\mu=Q/2$.

\begin{table}[t!]
\begin{center}
\footnotesize
\begin{tabular}{r | c c c c c c}
\hline\hline
$\mu_0=Q$ & NNLO (fb) & scale unc. (\%) & NNLL (fb) & scale unc. (\%)
& PDF unc. (\%) & PDF$+\as$ unc. (\%)
\\
\hline
$8\text{ TeV}$ & $9.92$ & $+9.3-1$0 & $10.8$ & $+5.4-5.9$ & $+5.6-6.0$ & $+9.3-9.2$\\
$13\text{ TeV}$ & $34.3$ & $+8.3-8.9$ & $36.8$ & $+5.1-6.0$ & $+4.0-4.3$ & $+7.7-7.5$\\
$14\text{ TeV}$ & $40.9$ & $+8.2-8.8$ & $43.7$ & $+5.1-6.0$ & $+3.8-4.0$ & $+7.5-7.3$\\
$33\text{ TeV}$ & $247$ & $+7.1-7.4$ & $259$ & $+5.0-6.1$ & $+2.2-2.8$ & $+6.1-6.1$\\
$100\text{ TeV}$ & $1660$ & $+6.8-7.1$ & $1723$ & $+5.2-6.1$ & $+2.1-3.0$ & $+5.7-5.8$\\
\hline\hline
$\mu_0=Q/2$ & NNLO (fb) & scale unc. (\%) & NNLL (fb) & scale unc. (\%)
& PDF unc. (\%) & PDF$+\as$ unc. (\%)
\\
\hline
$8\text{ TeV}$ & $10.8$ & $+5.7-8.5$ & $11.0$ & $+4.0-5.6$ & $+5.8-6.1$ & $+9.6-9.3$\\
$13\text{ TeV}$ & $37.2$ & $+5.5-7.6$ & $37.4$ & $+4.2-5.8$ & $+4.1-4.3$ & $+7.8-7.6$\\
$14\text{ TeV}$ & $44.2$ & $+5.5-7.6$ & $44.5$ & $+4.2-5.9$ & $+3.9-4.1$ & $+7.6-7.4$\\
$33\text{ TeV}$ & $264$ & $+5.3-6.6$ & $265$ & $+4.6-6.1$ & $+2.4-2.7$ & $+6.3-6.1$\\
$100\text{ TeV}$ & $1760$ & $+5.3-6.7$ & $1762$ & $+4.9-6.4$ & $+2.2-3.1$ & $+6.2-7.0$\\
\end{tabular}
\normalsize
\end{center}
\vspace{-0.5cm}
\caption{\label{tabla}\small
The total cross section and theoretical uncertainties for different center of mass energies, at NNLO and NNLL, for $\mu_0=Q$ and $\mu_0=Q/2$.
PDF and PDF$+\as$ uncertainties correspond to the resummed predictions, and are estimated using the sets of MSTW2008 at $90\%$ confidence level.
}
\end{table}

We focus now on the theoretical uncertainty arising from the missing higher order contributions, which is estimated by the scale variation indicated above.
In Table \ref{tabla} we present the total cross section predictions at NNLO and NNLL, together with the scale uncertainty.
We can observe that in all the cases the uncertainty of the resummed result is lower than the fixed order one. For instance, for $E_{cm}=14\text{ TeV}$ we find that the total uncertainty at NNLO is $17\%$ for $\mu_0=Q$, while it goes down to $11\%$ at NNLL.
This reduction is less important but still noticeable for $\mu_0=Q/2$, where it goes from $13\%$ to $10\%$ for the same center of mass energy.

Another interesting feature to notice is the stability of the resummed cross sections, being almost independent of the choice of the central scale.
The differences between the $\mu_0=Q$ and $\mu_0=Q/2$ results are below $2\%$ in the central value, and present very similar uncertainty bands.

In Table \ref{tabla} we also present the uncertainties coming from the strong coupling and parton flux determination for the resummed cross section.
These were estimated using the MSTW2008 $90\%$ C.L. error PDF sets \cite{Martin:2009bu}, which are known to provide very close results to the PDF4LHC working group recommendation for the envelope prescription \cite{Botje:2011sn}.
The results are very similar to the ones corresponding to the fixed order calculation \cite{deFlorian:2013jea}.
We can see that the PDF$+\as$ uncertainty is typically larger than the one arising from scale variation.

\begin{figure}[t!]
\begin{center}
\begin{tabular}{c c c}
\hspace{2.0cm}
\includegraphics[height=6cm]{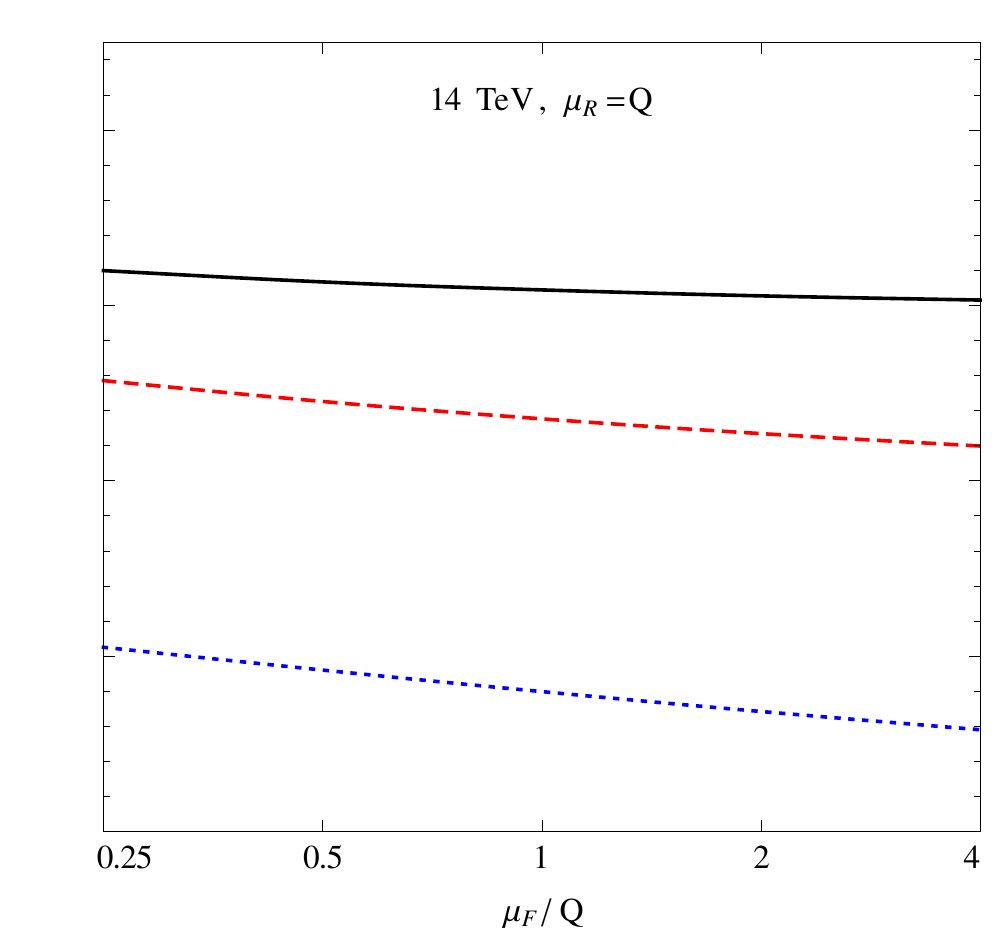}
&
\hspace{-12.75cm}
\includegraphics[height=6cm]{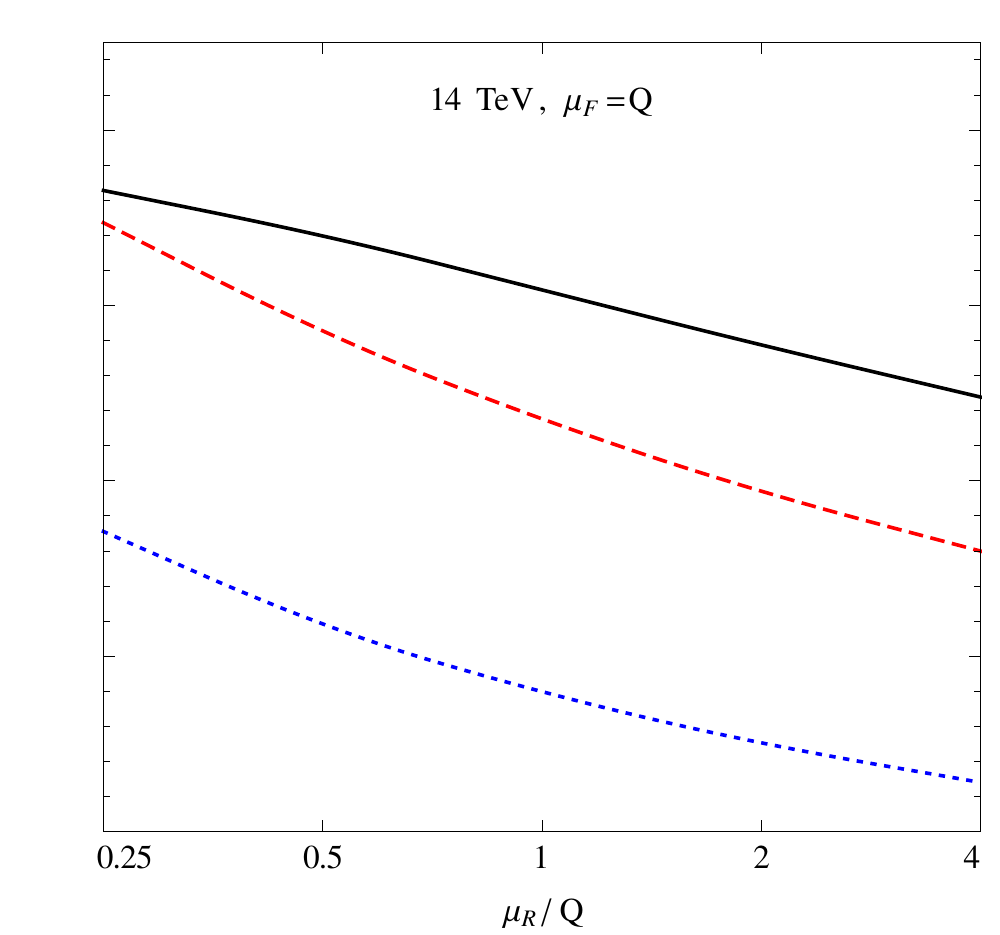}
&
\hspace{-12.75cm}
\includegraphics[height=6cm]{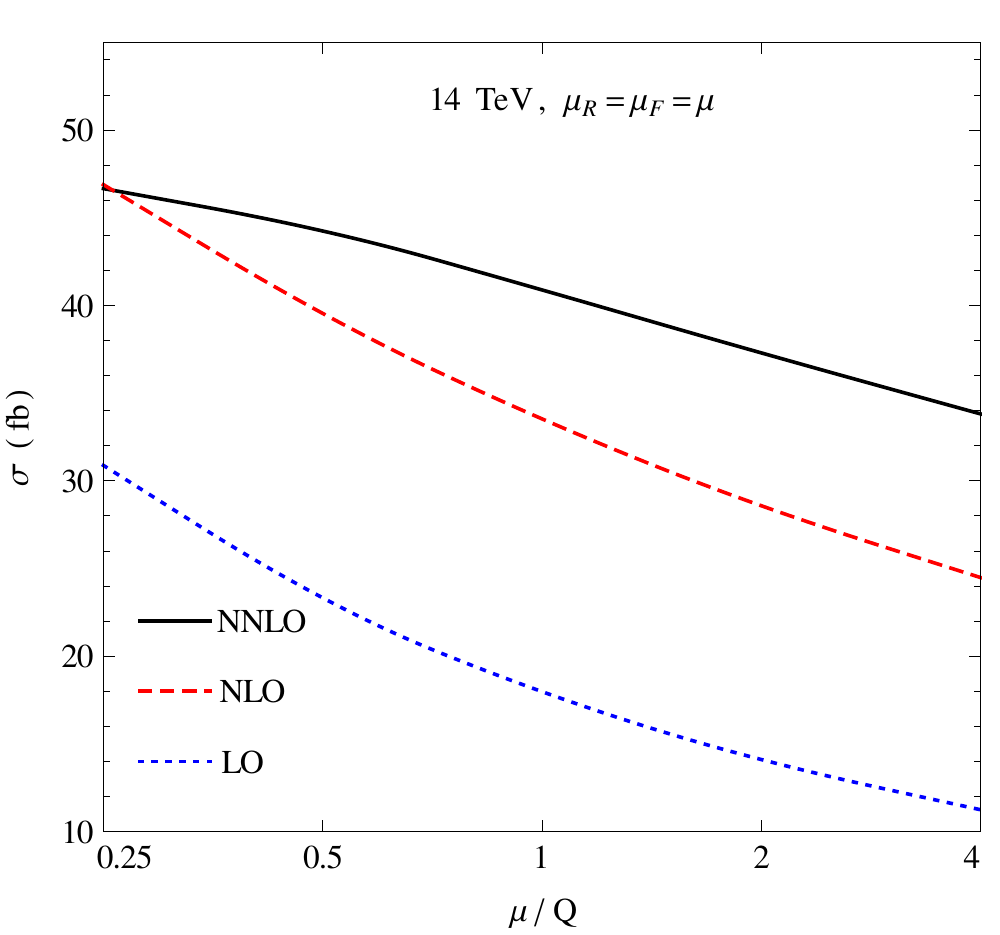}
\\
\hspace{2.0cm}
\includegraphics[height=6cm]{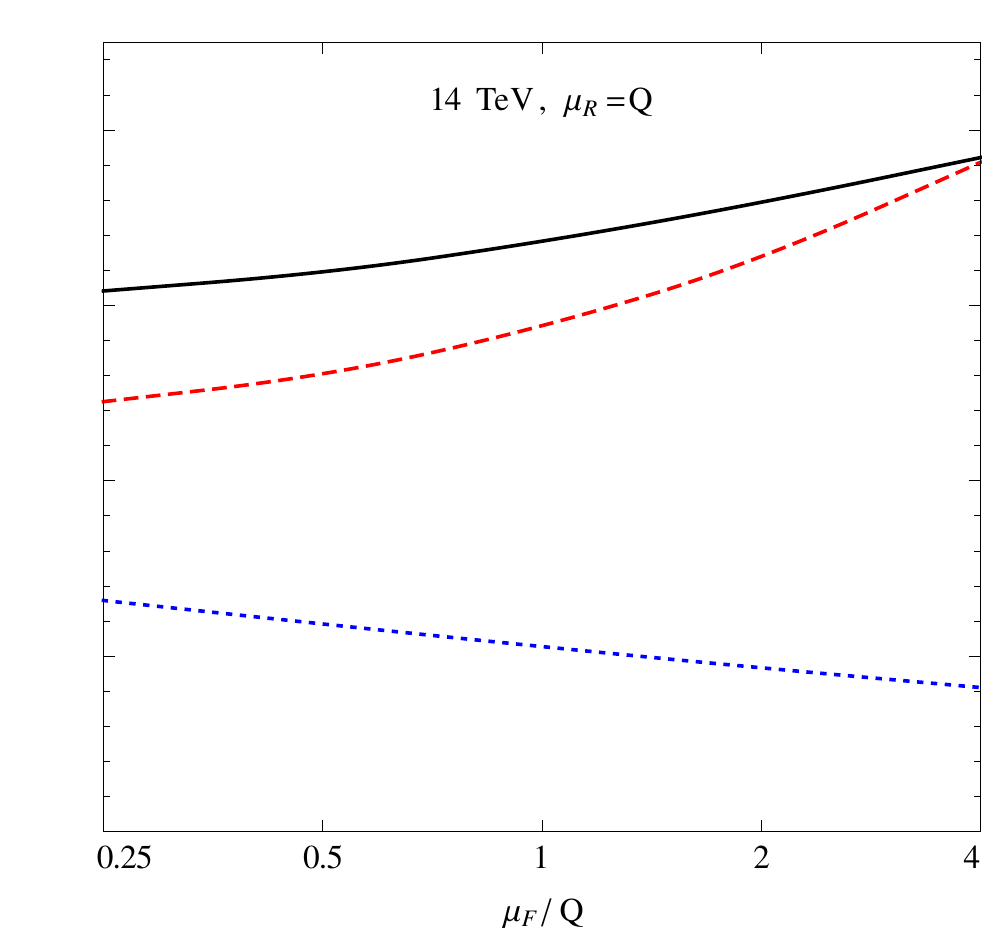}
&
\hspace{-12.75cm}
\includegraphics[height=6cm]{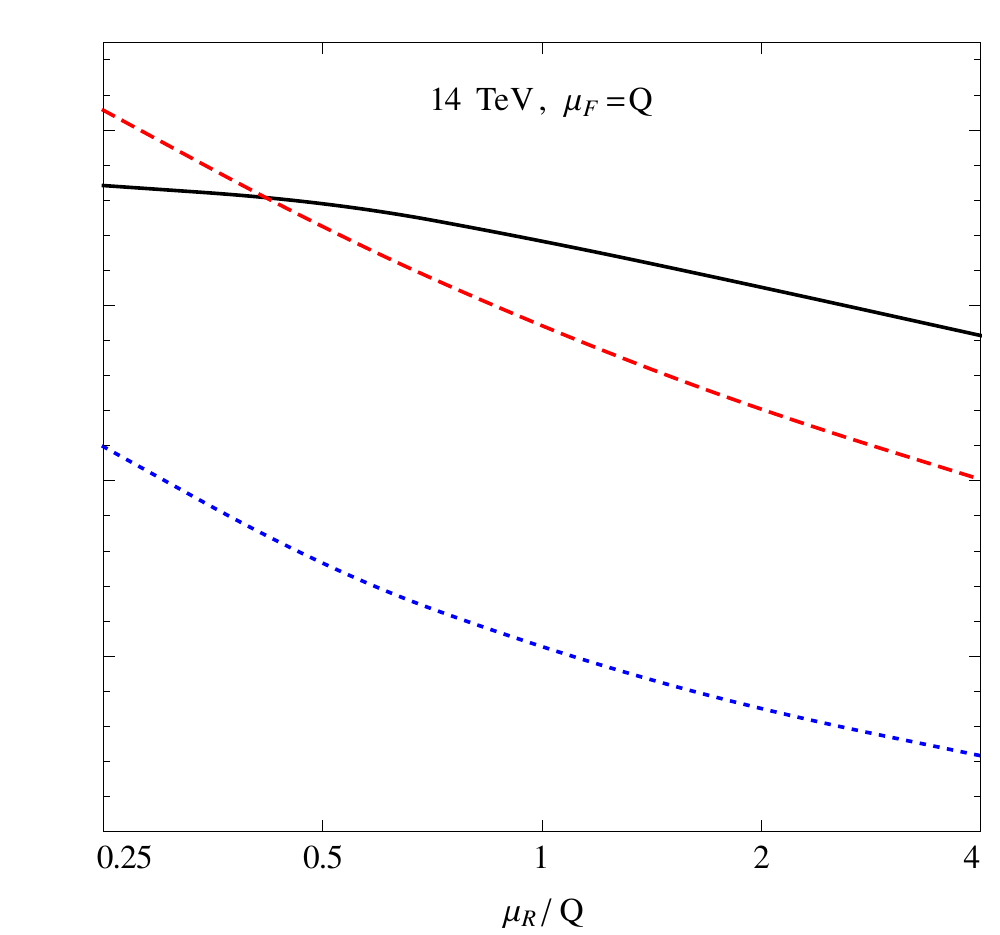}
&
\hspace{-12.75cm}
\includegraphics[height=6cm]{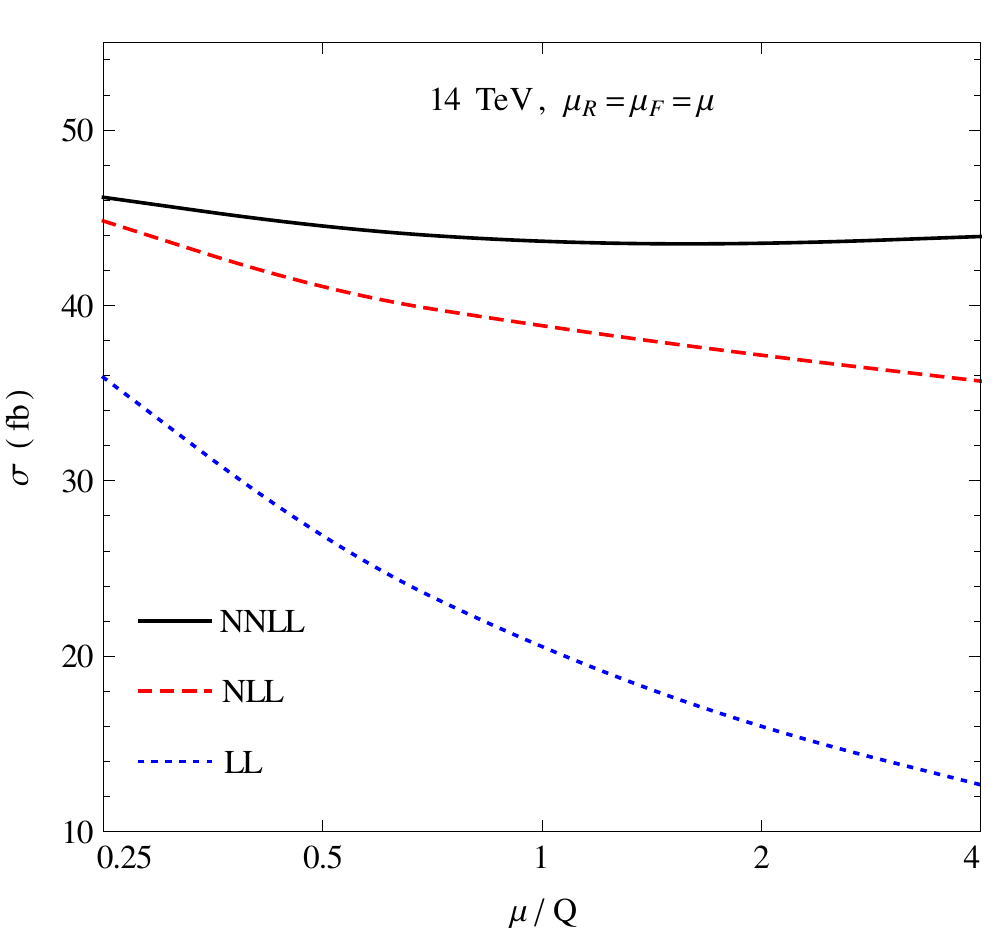}
\end{tabular}
\end{center}
\vspace{-0.7cm}
\caption{\small\label{scale14}
The scale dependence of the total cross section at $E_{cm}=14\text{ TeV}$, for the fixed order (upper) and resummed (lower) predictions.
The color coding is the same of Figure \ref{dsdQ_14_mu1}.
}
\end{figure}

In order to further illustrate the reduction of the uncertainty, we present in Figure \ref{scale14} the scale dependence of the total cross section, both for the fixed order and resummed predictions, at the different accuracy levels.
The plots in the left correspond to varying simultaneously the factorization and renormalization scales with $\mu_R=\mu_F=\mu$, the one in the center corresponds to the dependence on the renormalization scale for fixed $\mu_F=Q$, while the plot in the right shows the factorization scale dependence, for fixed $\mu_R=Q$.

In all the plots we can observe that the inclusion of higher order corrections reduces the scale dependence.
The contributions from resummation further reduce this dependence at NNLL, except for the $\mu_F$ dependence at fixed $\mu_R=Q$.
This last feature is also present for single Higgs production \cite{Catani:2003zt}, and suggests that the
rather flat dependence on $\mu_F$ at NNLO can be an accidental effect.

\begin{figure}
\begin{center}
\begin{tabular}{c c c c}
\hspace{0.0cm}
\includegraphics[height=7.89cm]{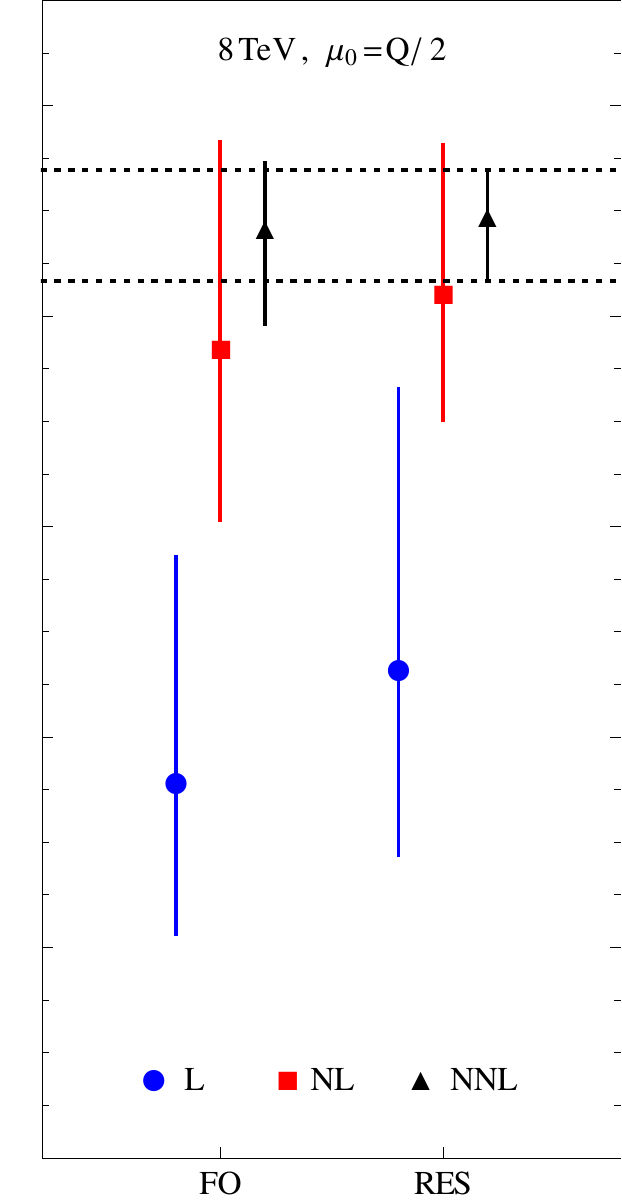}
&
\hspace{-8.87cm}
\includegraphics[height=7.89cm]{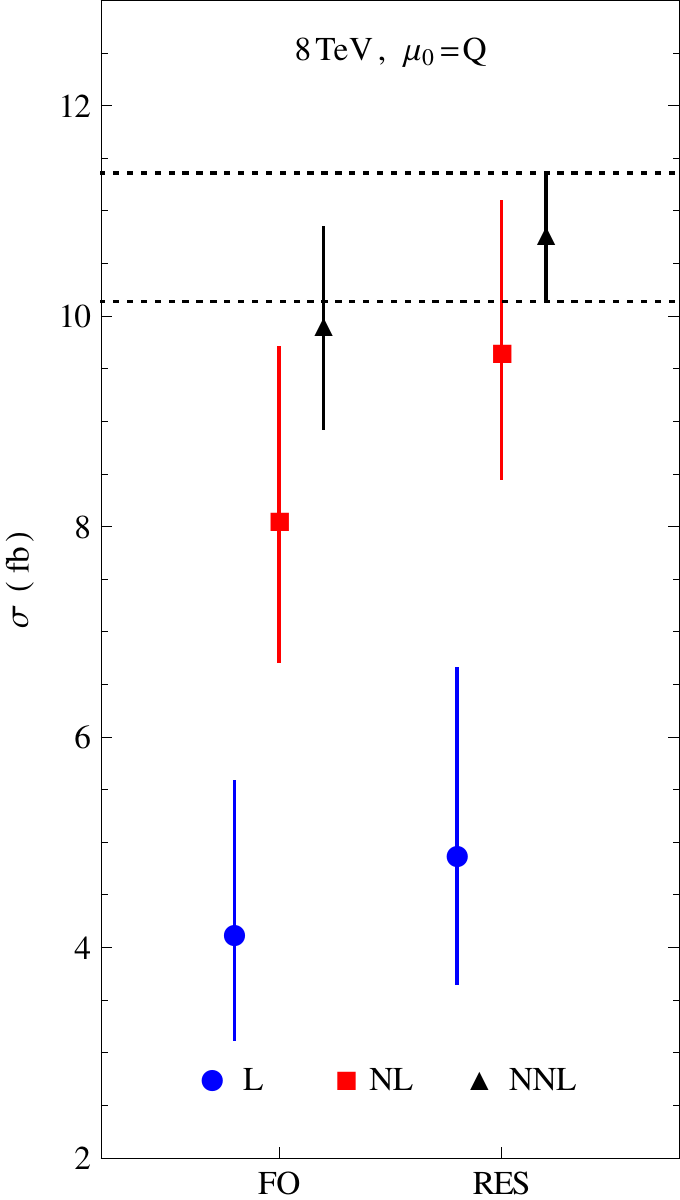}
&
\hspace{8.5cm}
\includegraphics[height=8cm]{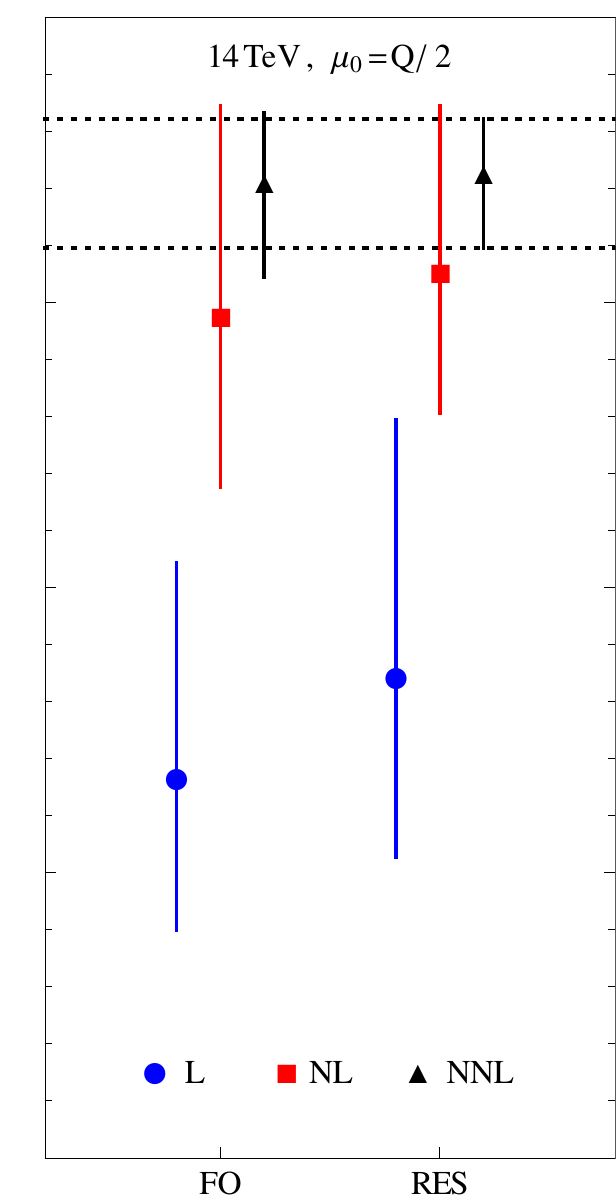}
&
\hspace{-8.9cm}
\includegraphics[height=8cm]{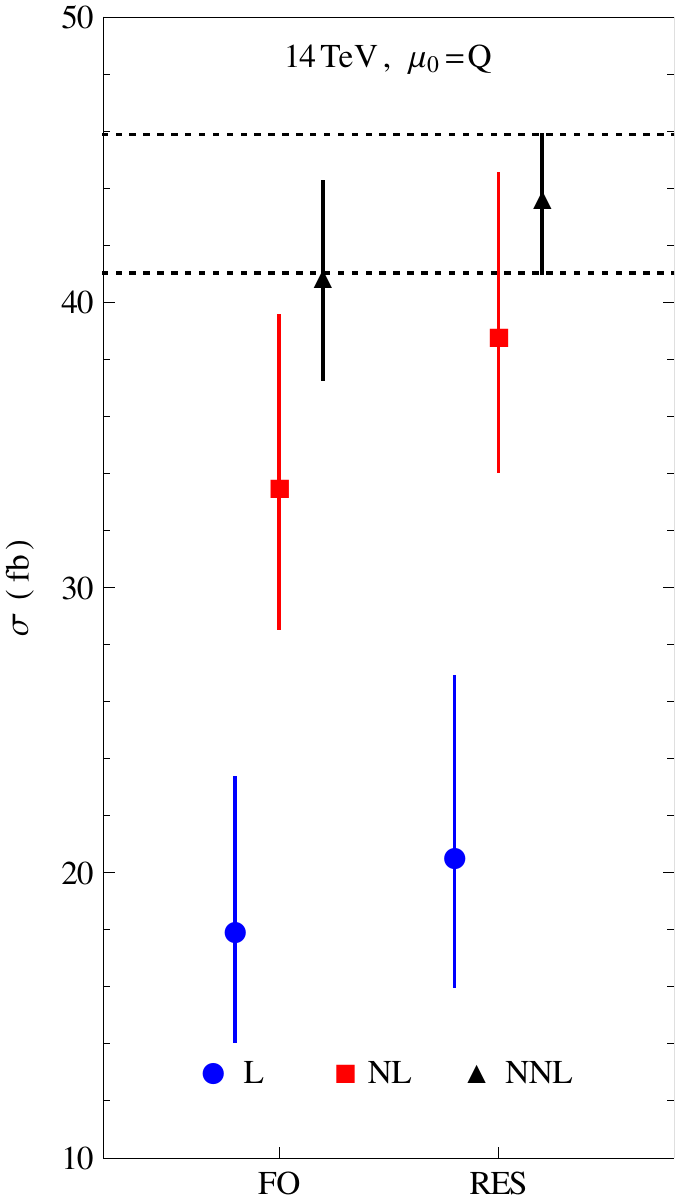}
\end{tabular}
\end{center}
\vspace{-0.7cm}
\caption{\label{total_8y14}\small
The total fixed order (FO) and resummed (RES) cross sections at leading (blue circle), next-to-leading (red square) and next-to-next-to-leading (black triangle) accuracy, for $E_{cm}=8\text{ TeV}$ (left) and $14\text{ TeV}$ (right), for both central scales $\mu_0=Q$ and $\mu_0=Q/2$.
The vertical solid lines indicate the scale uncertainty.
The horizontal dotted lines indicate in each case the best prediction (NNLL).
}
\end{figure}

We summarize our results for the total cross section in Figure \ref{total_8y14}, for $E_{cm}=8\text{ TeV}$ and $14\text{ TeV}$.
In this figure we present the value of the total cross section at LO, NLO and NNLO along with the LL, NLL and NNLL predictions.
The left panel of each plot shows the results corresponding to the central scale $\mu_0=Q$, while in the right one are those associated to $\mu_0=Q/2$.
The vertical lines indicate the scale uncertainty of each result.

The plots for both c.m. energies have similar features.
As was pointed out before, for $\mu_0=Q$ the corrections coming from threshold resummation are sizeable at every order, and the reduction of the total scale uncertainty at NNLL is notorious.
For $\mu_0=Q/2$ the increase on the total cross section is much smaller, though one can still observe a reduction of the uncertainty.
We can notice again the stability of the resummed prediction on the choice of the central scale.
This is illustrated in Figure \ref{total_8y14} by the horizontal dotted lines, which indicates the NNLL result for each value of $\mu_0$.
We can observe that the overlap between the two results is almost perfect, while in the case of the fixed order prediction there are much larger differences between them.

Finally, we comment on the results obtained in Ref. \cite{Shao:2013bz}. In that paper, a NLO$+$NNLL prediction for Higgs pair production was presented, based on the soft-collinear effective theory. That calculation did not include the matching to the NNLO cross section nor the coefficient $C_{HH}^{(2)}$, which were not available by the time of its publication.
They found that the remaining total scale uncertainty was below $8\%$, a value that is slightly below our estimation.
Also, their central value for the total cross section is below our result by about $2-3\%$.

\section{Conclusions}\label{conclusions}

We performe the soft-gluon threshold resummation up to NNLL accuracy, including consistently the matching to the NNLO cross section.
We work in the large top-mass approximation, normalizing our results by the exact LO dependence.

We find that the resummation results in an increase of the total cross section of $6.8\%$ for $E_{cm}=14\text{ TeV}$ and $\mu_0=Q$. The effect increases for lower center of mass energies and decreases for larger energies, as expected for the threshold contributions.
The increase in the total cross section, for the same value of c.m. energy, goes down to $0.65\%$ for $\mu_0=Q/2$.

The scale uncertainty is also reduced with respect to the fixed order prediction, going from $\pm8.5\%$ to $\pm5.5\%$.
The resummed prediction, including the corresponding uncertainty band, is found to be almost independent of the value chosen as the central scale, $\mu_0=Q$ or $\mu_0=Q/2$.
Given the similarity between the results, we can select in principle any of them as our final recommendation.
For the sake of definiteness, we recommend the usual setting $\mu_0=Q$ for the NNLL prediction, which for single Higgs production provides a result compatible with the recently computed N$^3$LO cross section \cite{Anastasiou:2015ema}.

The finite top-mass effects were analyzed at NLO in Ref. \cite{Grigo:2013rya} through the computation of subleading terms in the $1/M_t$ expansion, and in Ref. \cite{Maltoni:2014eza} via a reweighting technique that allows to exactly include the one- and two-loop amplitudes. Based on those studies, we can estimate the uncertainties coming from the use of the effective theory to be $\sim\pm10\%$ for the total cross section.
The uncertainty coming from the missing higher orders of the QCD perturbative expansion is now definitely below that value, and also below the strong coupling and parton flux determination uncertainties, which is about $\pm 7.5\%$ for $14\text{ TeV}$.

\subsection*{Acknowledgments}

\vspace*{-2mm}
\noindent
This work has been partially supported by the European Union through contract
PITN-GA-2010-264564 ({\it LHCPhenoNet$\,$})
and by UBACYT, ANPCYT and Conicet.

\vspace*{-3mm}

\bibliography{biblio}

\end{document}